\documentclass[journal]{IEEEtran}

\usepackage[pdftex]{graphicx}
\usepackage{mathrsfs,amsmath}
\usepackage{amssymb}
\usepackage[hyphens]{url}
\usepackage[export]{adjustbox}
\usepackage{comment}
\graphicspath{{images/}}
\begin{document}
	\title{Mitigating Aberration-Induced Noise: A Deep Learning-Based Aberration-to-Aberration Approach}
	
	\author{Mostafa~Sharifzadeh,
		Sobhan Goudarzi,
		An Tang,
		Habib~Benali,~\IEEEmembership{Fellow,~IEEE,}
		and~Hassan~Rivaz,~\IEEEmembership{Senior~Member,~IEEE}
		\thanks{M. Sharifzadeh, H. Benali, and H. Rivaz are with the Department of Electrical and Computer Engineering, Concordia University, Montreal, QC, Canada. S. Goudarzi is with the Physical Science Platform, Sunnybrook Research Institute, Toronto, ON, Canada. A. Tang is with the Department of Radiology, Radiation Oncology and Nuclear Medicine, Université de Montréal, Montreal, QC, Canada.}
		
		\thanks{This article has been accepted for publication in IEEE Transactions on Medical Imaging. Citation information: DOI 10.1109/TMI.2024.3422027
		\\© 2024 IEEE. Personal use is permitted, but republication/redistribution requires IEEE permission.}
	}
	\maketitle
	
	\begin{abstract}
		One of the primary sources of suboptimal image quality in ultrasound imaging is phase aberration. It is caused by spatial changes in sound speed over a heterogeneous medium, which disturbs the transmitted waves and prevents coherent summation of echo signals. Obtaining non-aberrated ground truths in real-world scenarios can be extremely challenging, if not impossible. This challenge hinders the performance of deep learning-based techniques due to the domain shift between simulated and experimental data. Here, for the first time, we propose a deep learning-based method that does not require ground truth to correct the phase aberration problem and, as such, can be directly trained on real data. We train a network wherein both the input and target output are randomly aberrated radio frequency (RF) data. Moreover, we demonstrate that a conventional loss function such as mean square error is inadequate for training such a network to achieve optimal performance. Instead, we propose an adaptive mixed loss function that employs both B-mode and RF data, resulting in more efficient convergence and enhanced performance. Finally, we publicly release our dataset, comprising over 180,000 aberrated single plane-wave images (RF data), wherein phase aberrations are modeled as near-field phase screens. Although not utilized in the proposed method, each aberrated image is paired with its corresponding aberration profile and the non-aberrated version, aiming to mitigate the data scarcity problem in developing deep learning-based techniques for phase aberration correction.
		Source code and trained model are also available along with the dataset at \url{http://code.sonography.ai/main-aaa}.
	\end{abstract}
	
	\begin{IEEEkeywords}
		phase aberration, ultrasound imaging, adaptive mixed loss, neural networks.
	\end{IEEEkeywords}
	
	\section{Introduction}
	\label{sec:Introduction}
	Ultrasound imaging is commonly used for medical diagnosis and image-guided interventions \cite{Masoumi2023} due to its advantages, such as portability, non-invasiveness, high temporal resolution, and low cost. However, it often suffers from artifacts, with phase aberration as one of the main sources of image quality degradation \cite{Pinton2011}.
	
	Spatially varying sound speed while traveling through a heterogeneous medium is the origin of the phase aberration effect \cite{Flax1988}. In a perfect homogeneous medium, where the sound speed is constant and known, the traveling time for a pulse from any transducer element to any point in the medium can be calculated using simple geometry. Therefore, the required time delays that need to be applied to each element can be determined accurately to compensate for traveling path length differences and form the desired beam in transmit beamforming. Similarly, in receive beamforming, time delays can be calculated and applied to received echo signals in order to sum them coherently. In practice, however, the human body is a heterogeneous medium, where, for instance, the sound speed in fat and muscle is approximately 1460 m/s and 1610 m/s, respectively, which indicates a variation of almost 10\%  \cite{Goss1980}. The variation is even higher in applications such as transcranial imaging \cite{OReilly2013}, where the average sound speed in the skull is nearly 2740 m/s \cite{Goss1978} and can vary, reaching more than 4000 m/s \cite{Peterson2003}. The phase aberration effect in a heterogeneous medium alters the focal point in focused imaging and perturbs the flat wavefront propagation in plane-wave imaging during the transmission, and prevents coherent summation of echo signals in both imaging techniques during the reception, all of which cause suboptimal image quality.
	
	\subsection{Related Work}
	\label{ssec:Introduction:RelatedWork}
	Aberration correction has been studied for years in the medical ultrasound community, as it can improve anatomical fidelity and spatial localization, both of which lead to improved diagnostic accuracy and precision of image-guided interventions.
	Several techniques attempted to estimate delay errors by maximizing the cross-correlation \cite{Flax1988} or minimizing the absolute differences between RF signals received at adjacent array elements \cite{Karaman1993}, maximizing mean speckle brightness in a region of interest \cite{Nock1989}, or incorporating a virtual point reflector generated by iterative time reversal focusing \cite{Osmanski2012}. Li \textit{et al.} utilized the generalized coherence factor for reducing focusing errors, especially the ones caused by sound speed inhomogeneities \cite{Li2003}. Napolitano \textit{et al.} analyzed lateral spatial frequency content in reconstructed images to find the optimal sound speed for subsequent imaging that maximizes the focus quality \cite{Napolitano2006}. Shin \textit{et al.} employed an adaptive filtering technique called frequency-space prediction filtering (FXPF), which presupposes the existence of an autoregressive model across the echo signals received at the transducer elements and removes any components that do not conform to the established model \cite{Shin2017, Shin2018a}.
	
	As opposed to methods that model the phase aberration effect by a fixed near-field phase screen in front of the transducer, the locally adaptive phase aberration correction technique \cite{Chau2019} assumed a spatially varying near-field phase screen and employed multistatic synthetic aperture data to perform the correction at each point adaptively. Lambert \textit{et al.} suggested compensating for the spatially-distributed aberrations by decoupling aberrations undergone by the outgoing and incoming waves utilizing the distortion matrix built from the focused reflection matrix, which contains the responses between virtual transducers synthesized from the transmitted and received focal spots \cite{Lambert2020, Lambert2021, Lambert2021a}.
	
	A different category of techniques utilizes echo signals as input and returns an estimation of the spatial distribution of sound speed in a given medium \cite{Sanabria2018a, Raua}. Although these methods are not an immediate approach for aberration correction, the estimated distribution can be subsequently employed to compensate for the phase aberration effect, for instance, by reconstructing the image by computing beamforming delays assuming that sound travels on straight line paths \cite{Jaeger2015} or using a set of refraction-corrected delays based on the Eikonal equation \cite{Ali2018}, which can be efficiently solved using the fast marching method \cite{Sethian1999}.
	The computed ultrasound tomography in echo mode (CUTE) method correlated the phase shifts across a sequence of beamformed plane-wave images obtained with different steering angles and exploited that to estimate the distribution of sound speed \cite{Stahli2020}. Jakovljevic \textit{et al.} proposed and solved a model via gradient descent that establishes a connection between the local speed of sound along a wave propagation path and the average speed of sound over that path \cite{Jakovljevic2018}, where the latter is measured using the method proposed in \cite{Anderson1998}. Although the efficacy of this model was demonstrated in layered heterogeneous media, the performance often drops when the variations of sound speed are not insignificant along the lateral axis. Rehman \textit{et al.} introduced a tomography-based method that directly accounts for propagation paths between the scattering volume and each transducer element to mitigate that issue \cite{Ali2019}. They also proposed an inverse-modeled phase aberration computed tomography (IMPACT) framework, which utilizes multistatic synthetic aperture data, estimates the global average sound speed \cite{Brevett2022} by maximizing coherence for each point, applies an inversion to compute the local sound speed, and finally exploits them in two different Eikonal equation-based and wavefield correlation-based distributed aberration correction techniques \cite {Ali2022}. In addition to approaches that aim to rectify the aberrated image, beamformers exist which are designed to be robust to this artifact by incorporating singular value decomposition of a defined matrix comprising backscattered data for multiple plane-wave transmissions \cite{Bendjador2020}.
	
	Recently, utilizing deep learning (DL)-based techniques for phase aberration correction has attracted growing interest. Among one of the first endeavors, Sharifzadeh \textit{et al.} assumed a near-field phase screen model and demonstrated that a convolutional neural network (CNN) could estimate delay errors, or the aberration profile, directly from the B-mode image with a high accuracy \cite{Sharifzadeh2020}. Feigin \textit{et al.} simulated a dataset using the k-Wave toolbox \cite{Treeby2010}, wherein the organs in tissue were modeled as uniform ellipses over a homogeneous background with different sound speeds. They trained a CNN on the dataset to estimate sound speed distribution, taking raw RF channel data of three plane-wave transmissions as inputs \cite{Feigin2020a}. In a similar approach, demodulated in-phase and quadrature (IQ) data were provided to the network as the inputs \cite{KhunJush2021}. Additional comparable methodologies have been proposed in the literature \cite{Jush2020, Young2022} for the same purpose. Koike \textit{et al.} trained a network by mapping aberrated RF inputs to their corresponding aberration-free RF target outputs \cite{Koike2023}. Shen \textit{et al.} utilized a CNN to estimate the aberrated point spread function from beamformed IQ data and subsequently applied the inverse filter to rectify the data \cite{Shen2022}.
	Additionally, there are DL-based beamformers designed to exhibit robustness to the aberration by suppressing off-axis scattering \cite{Luchies2020} or by mapping images beamformed with randomly perturbed sound speed values to clean images beamformed with a reference sound speed value \cite{Khan2022}.
	
	\subsection{Contributions}
	\label{ssec:Introduction:Contributions}
	The main challenge in utilizing DL-based approaches for correcting phase aberration is the absence of a reliable ground truth. As a result, the aforementioned methods had to rely solely on simulated data for training, leading to a drop in performance when testing on experimental data due to the domain shift problem \cite{Sharifzadeh2021b}. Recent studies have recognized the need to eliminate the requirement of ground truths; however, even in such efforts, reconstructed images with a fixed sound speed value of 1540 m/s were still considered clean images \cite{Khan2022}. Our contributions can be summarized as follows:
	
	\begin{enumerate}
		\item We propose the first DL-based aberration correction method that eliminates the need for ground truth in the training phase. Both input and target output are randomly aberrated RF data, which enables us to use real experimental data for training, fine-tuning, or both, without any explicit assumption regarding the presence or absence of phase aberrations.
		\item Our training setup presents a significant challenge as both the input and desired output of the network contain aberrations that randomly differ in each frame and epoch. Adding to the complexity is the fact that RF data includes high-frequency components. We demonstrate that a conventional loss function such as mean square error (MSE) is inadequate for training such a network. To address this challenge, we introduce a loss function that incorporates both B-mode and RF data and evaluate its performance.	
		\item We publicly release a dataset comprising 1,802 sets of single plane-wave images (RF data). Each set includes 100 aberrated versions of the same realization. Although not utilized in the proposed method of this study, corresponding aberration profiles and non-aberrated versions are also included for comprehensiveness. To the best of our knowledge, this is the first dataset practically suitable for developing DL-based techniques in this domain, given its size and structure.
		\item We release our code and trained model online. Additionally due to the unavailability of a publicly available implementation of the FXPF method, which is compared with the proposed method, we make our own implementation publicly accessible.
	\end{enumerate}
	The proposed method \textbf{m}itigates \textbf{a}berration-\textbf{i}nduced \textbf{n}oise using an \textbf{a}berration-to-\textbf{a}berration \textbf{a}pproach, which we name MAIN-AAA, and show that it substantially improves aberrated images in simulation and phantom experiments. As one of the authors (AT) is a radiologist, we were also able to visually corroborate improvements in \textit{in-vivo} images.
	\section{Methodology}
	\label{sec:Methodology}
	\subsection{Aberration-to-Aberration Approach}
	\label{ssec:A2A}
	Let us consider a set of noisy measurements denoted by $a = (a_1, a_2, ..., a_N)$, representing the recorded signal amplitude corresponding to reflection from a particular point within a medium.
	To estimate the true amplitude, a common approach involves finding a value $\hat{a}$ that minimizes the expected deviation from measurements according to a loss function $L$:
	\begin{equation}
		\arg\min_{\hat{a}} \mathbb{E}_{a}{\{L(\hat{a}, a)\}}.
	\end{equation}
	For $L(\hat{a}, a) = (\hat{a} - a)^2$, it is straightforward to demonstrate that this minimum occurs at the arithmetic mean of the measurements.
	Training neural networks as regression models is a generalization of this point estimation approach, which means that training a network with infinite samples utilizing an MSE loss function estimates the expectation of the target samples \cite{Lehtinen2018}.
	
	In the context of ultrasound images, tissue response can be represented as a single point within a high-dimensional manifold. Phase aberrations and artifacts, such as those caused by sidelobes and multiple scattering, can shift this point, deviating from its original position. However, these artifacts tend to be inconsistent across different images, whereas the tissue response remains consistent. Consequently, when training a network using randomly aberrated images, the objective is to disentangle these artifacts from the tissue response by interrogating different aberrated instances.
	\subsection{Phase Aberration Model}
	\label{ssec:PhaseAberrationModel}
	We modeled the phase aberration effect by assuming a near-field phase screen in front of the transducer, which introduces different delay errors to each transducer element during both transmission and reception. Although this model does not make any assumptions regarding the spatial distribution of sound speed within the medium, it proves particularly useful in scenarios where an aberrator layer in front of the transducer is so dominant that other sources of aberration in the remainder of the medium are negligible. An example is imaging overweight subjects, where the wave must propagate through a thick layer of fat found in the near-field \cite{Shin2018a}. In these cases, slight lateral variations in the thickness of the strong aberrator layer may impose strong aberrations, often impeding the optimal performance of methods designed to estimate the distribution of sound speed \cite{Ali2019}.
	
	The aberration profile in this model is represented as an array, where each element of the array corresponds to a delay error value assigned to a specific transducer element. Aberration profiles are characterized by their strength and correlation length. The strength is defined as its root mean square amplitude in nanoseconds, and the correlation length, which represents the spatial frequency content, is defined as the full width at half maximum of its autocorrelation in millimeters \cite{Dahl2005a}. An aberration profile becomes stronger and induces more degradation effect as its strength is increased, and its correlation length is decreased, which means higher amplitude with more fluctuation across the aperture \cite{Sharifzadeh2020}. Experiments were conducted in some literature to estimate the parameters of aberration profiles. For instance, the strength and correlation length for \textit{in-vivo} and \textit{ex-vivo} breast tissue are reported as 28.0 ns, 3.48 mm, and 66.8 ns, 4.3 mm respectively \cite{Hinkelman1995, Fernandez2001}. We generated random aberration profiles by convolving a Gaussian function with Gaussian random numbers \cite{Dahl2005a}, where they were varied uniformly in strength and correlation length ranging from 20 to 80 ns, and from 4 to 9 mm, respectively, to encompass a broad range of tissues.
	
	\subsection{Phase Aberration Implementation}
	We first explain the methodology used for introducing phase aberration into simulated and experimental phantom data. Details regarding the simulation or acquisition of data, as well as where each approach was employed, will be discussed later. The Supplementary Video, at \textit{http://code.sonography.ai/main-aaa}, also provides an overview.
	
	\subsubsection{Simulated Aberration}
	\label{sec:Methodology:PhaseAberrationImplementation:SimulatedAberration}
\begin{figure}
		\centering
		\includegraphics[width=0.9999\linewidth]{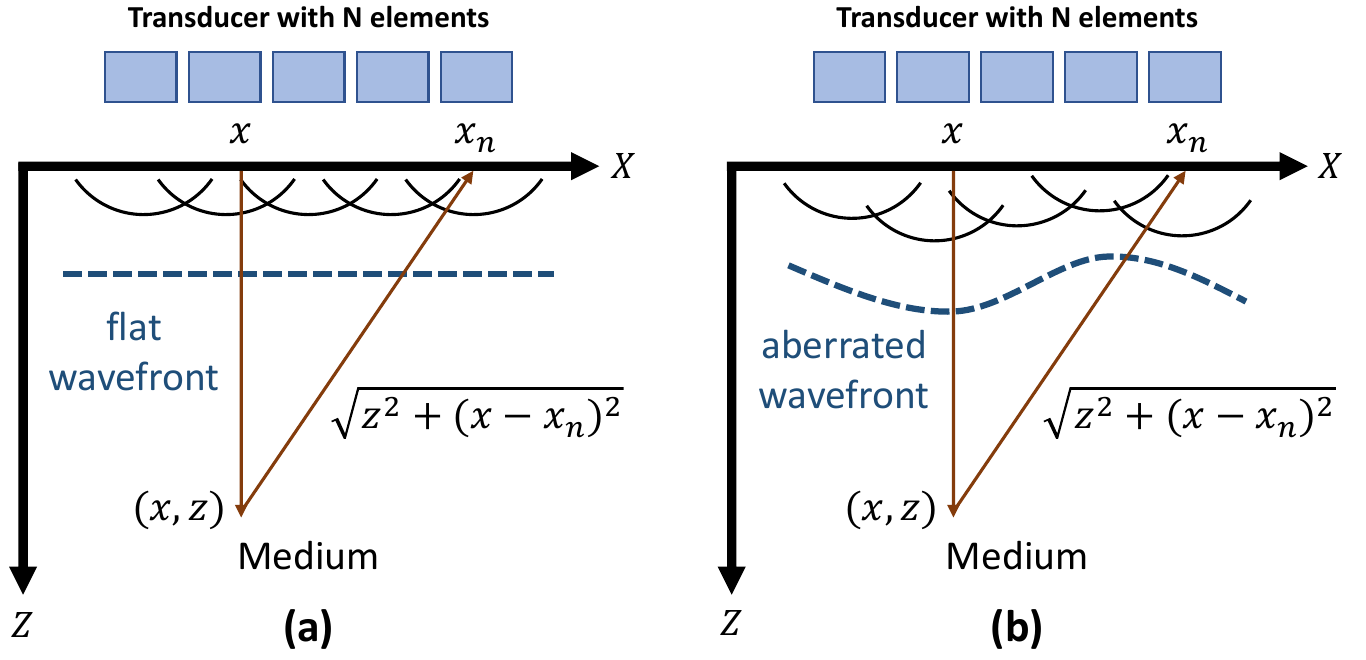}
		\caption{A typical configuration of ultrasonic imaging systems in the (a) absence and (b) presence of a near-field phase screen.}
		\label{fig-aberration-implementation}
		\vspace{-15pt}
	\end{figure}

	To introduce the aberration effect into simulated data, we utilized full synthetic aperture data and synthesized aberrated plane-wave images under linear and steady conditions. Fig. \ref{fig-aberration-implementation} demonstrates a typical configuration of ultrasonic imaging systems in the (a) absence and (b) presence of a near-field phase screen, which can be defined by an aberration profile $\tau_a$. A linear array transducer consisting of $N$ elements is positioned in direct contact with the imaging medium of interest. The array is oriented such that the x-axis is parallel to its length, while the depth direction within the imaging medium is represented by the z-axis.
	After a single plane-wave transmission, the received echo signal at time $t$ by element $n$ located at $x_n$ can be calculated using full synthetic aperture data as follows:
	\begin{flalign}
		\label{eq-RF}
		RF\left(x_n,t\right)=\sum_{m=1}^{N}{RF_{fsa}\left(x_m,x_n,t+\tau_a\left(x_m\right)\right)},
	\end{flalign}
	where $RF_{fsa}(x_m,x_n,t)$ is the received echo signal at time $t$ by element $n$ located at $x_n$ solely due to excitation of element $m$ located at $x_m$ and $\tau_a\left(x_m\right)$ is the delay error that element $m$ experiences according to the aberration profile $\tau_a$.
	In the absence of aberration, as shown in Fig. \ref{fig-aberration-implementation}(a), we can assume synchronous excitation times for all piezoelectric elements during synthesizing, equivalent to transmitting a flat wavefront. In this case, the delay error $\tau_a$ equals zero for all transducer elements. However, to simulate the phase aberration effect during transmission, as shown in Fig. \ref{fig-aberration-implementation}(b), we assumed asynchronous excitation times for piezoelectric elements by applying delay errors imposed by the aberration profile.
	In the absence of phase aberration, the required time for the acoustic wave to travel to point $(x, z)$ and return to the transducer element $n$ located at $x_n$ is
	\begin{flalign}
		\tau(x_n,x,z)=(z+\sqrt{z^2+{(x-x_n)}^2})/c,
	\end{flalign}
	where $c$ is the sound speed.
	The phase aberration effect in reception was implemented as a set of time delay errors corresponding to backscattered signals and according to the aberration profile $\tau_a$. To this end, and given the calculated time delay, each point $(x,z)$ within the region of interest can be reconstructed as
	\begin{flalign}
		\label{eq-s}
		s(x,z)=\sum_{n=k-[a/2]}^{k+[a/2]}{RF(x_n,\tau(x_n,x,z)+\tau_a(x_n))},
	\end{flalign}
	where $k$ is the nearest transducer element to $x$, and [.] represents rounding to the nearest integer. Aperture size $a$ determines the number of elements that contribute to the signal and can be expressed using the $f\text{-}$number, which was set to 1.75 in this study and is defined as $F=z/a$.
	In summary, delay errors $\tau_a(x_m)$ and $\tau_a(x_n)$ in equations (\ref{eq-RF}) and (\ref{eq-s}) contribute to the aberrations that occur during transmission and reception, respectively, where the former simulates asynchronous excitation of piezoelectric elements during synthesizing the plane-wave and the latter disorders time delays corresponding to received echo signals.
	
	\subsubsection{Quasi-physical Aberration}
	\label{sec:Methodology:PhaseAberrationImplementation:NumericalAberration}
	Our approach for introducing a quasi-physical aberration to an experimental phantom required programming a Vantage 256 research scanner. We programmed the scanner to excite transducer elements asynchronously according to a given aberration profile. To this end, delay errors corresponding to each element were calculated in wavelengths of the transducer center frequency and written to the \textit{TX.Delay} array, provided by the scanner programming interface. resulting in the generation of an aberrated wavefront during single plane-wave imaging. Moreover, delay errors introduced by the aberration profile were taken into account during the reception process for reconstructing the image.
	
	\subsubsection{Physical Aberration}
	To introduce a physical aberration to the experimental phantom, we placed an uneven layer of chicken bologna between the probe and the phantom, where the thickness of the left and right halves was approximately 3 mm and 6 mm, respectively. Although the precise value of the sound speed within this layer was unknown, we could be confident that it introduced the aberration effect due to its uneven thickness and observing the effect in the resulting image. To ensure proper contact, we filled the gap between the thinner half and the probe's surface with conductive gel and positioned the center of the probe at the discontinuity.
	
	\subsection{Datasets}
	\label{ssec:Datasets}
	\subsubsection{Simulated}
	\label{ssec:Datasets:Simulated}
\begin{figure}
		\centering
		\includegraphics[width=0.9999\linewidth]{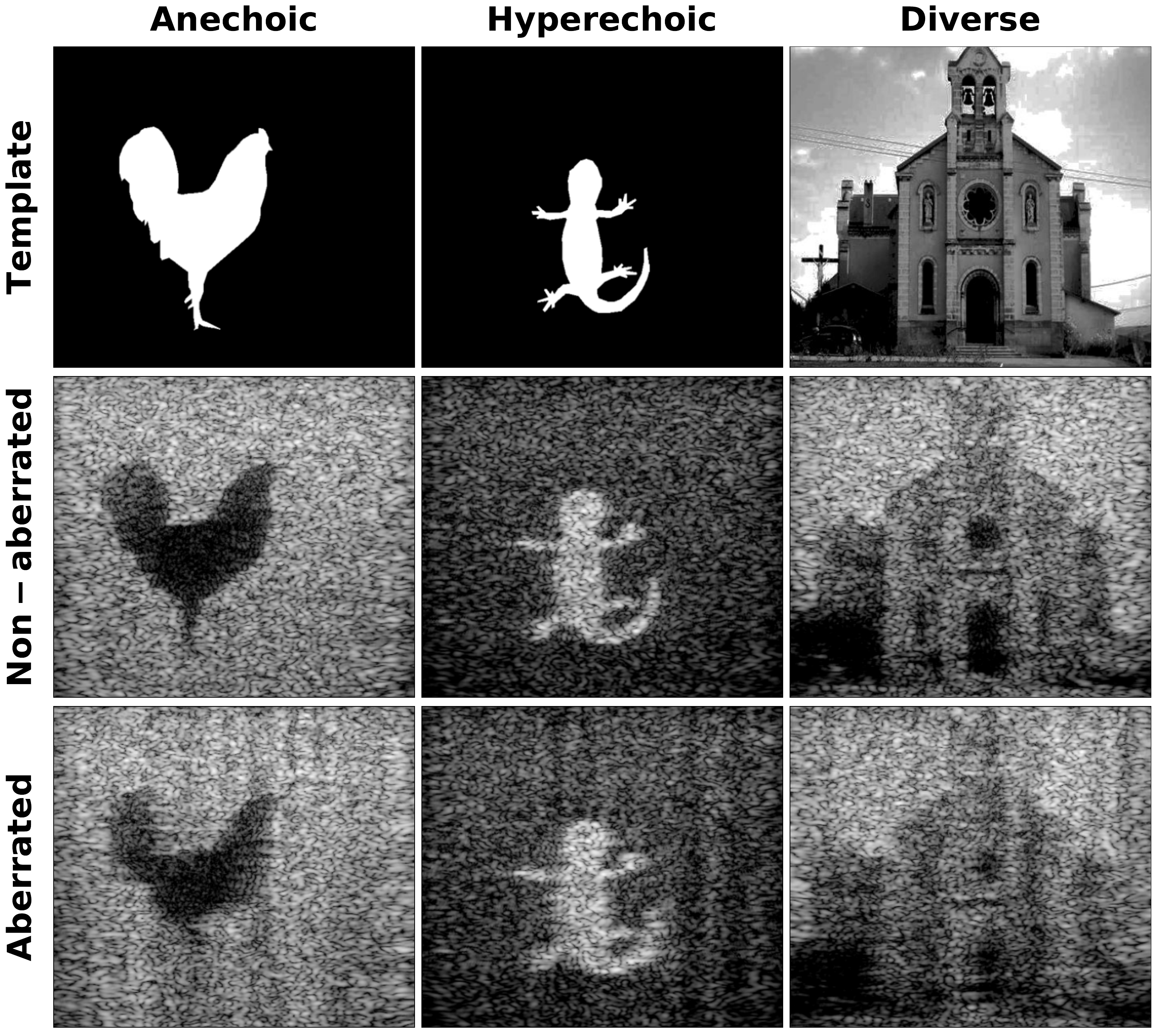}
		\caption{Samples from the simulated dataset. The left and middle columns showcase examples of anechoic and hyperechoic regions generated using arbitrary segmentation masks. The right column presents a diverse echogenicity example generated based on a natural image. For each case, the template, non-aberrated, and a sample aberrated version are presented in the first to third rows, respectively. Templates and non-aberrated images are included solely for visualization purposes and were not utilized in the proposed method.}
		\label{fig-simulated-dataset-samples}
		\vspace{-18pt}
	\end{figure}

	We simulated a synthetic dataset consisting of 1802 image sets using the publicly available Field II simulation package \cite{Jensen1992, Jensen1996}, containing an average scatterer density of 60 per resolution cell (fully developed speckle pattern). The scatterers were uniformly distributed inside a phantom measuring 45 mm in the lateral and 40 mm in the axial direction, positioned at an axial depth of 10 mm from the face of the transducer.
	We introduced contrast to the images by incorporating five different types of echogenicities: anechoic regions, hypoechoic regions, hyperechoic regions, diverse echogenicities, and point targets.
	To generate the first three types, we took 600 samples (200 samples per type) from a publicly available dataset, known as XPIE \cite{Xia2017}, which included segmented natural images. We then disregarded natural images and resampled only their corresponding segmentation masks to match the phantom's dimensions.
	Finally, the amplitude of scatterers located inside the mask was multiplied by a weight, which was zero for anechoic regions, a uniform random number $\in [0.063, 0.501]$ for -12 dB to -3 dB hypoechoic regions, and a uniform random number $\in [2, 15.8]$ for +3 dB to +12 dB hyperechoic regions.
	To enrich the range of echogenicity, we obtained an additional 1000 samples from the XPIE dataset, but this time we discarded the segmentation masks and instead resampled only the natural images with the same dimensions as the phantom. These images were then converted to grayscale, and similar to \cite{Hyun2019a}, the pixel intensities were utilized to weight the scatterers' amplitude according to their respective positions via bilinear interpolation. To enhance the contrast of the ultrasound images, we preprocessed natural images by performing histogram equalization and thresholding pixel values below 0.1 to zero and those above 0.9 to 1. Leveraging natural images and masks for simulation, as described in this subsection, offers the advantage of providing the network with a broader range of features compared to images containing only cysts or selectively chosen shaped regions.
		To simulate the remaining 200 sets, we introduced multiple randomly positioned point targets to each, where the number of them was determined by a uniform random number $\in [10, 20]$, and their amplitudes were set randomly by drawing from a uniform distribution between 12 dB to 16 dB higher than the mean amplitude of other scatterers.
		In addition, two test sets were simulated for evaluation purposes: a contrast test set and a resolution test set. The former comprised two anechoic cysts with diameters of 10 mm and 15 mm at central lateral positions and depths of 10 mm and 28 mm, respectively. The latter included a total of 19 point targets arranged in a vertical line at the central lateral position and two horizontal lines at depths of 10 mm and 30 mm.
	\noindent The transducer settings used for simulation were similar to those of the 128-element linear array L11-5v (Verasonics, Kirkland, WA) and are summarized in Table~\ref{tbl-prob-settings}.
	\begin{table}[h]
		\caption{The settings of linear array transducer L11-5v}
		\label{tbl-prob-settings}
		\setlength{\tabcolsep}{8pt}
		\def\arraystretch{1}%
		\begin{tabular}{p{115pt}p{50pt}p{40pt}}
			\hline
			\vspace{2pt}
			\textbf{Parameter}&
			\vspace{2pt}
			\textbf{Value}&
			\vspace{2pt}
			\textbf{Unit}
			\vspace{2pt}\\
			\hline
			\vspace{0pt}
			Number of Elements& 
			\vspace{0pt}
			128&
			\vspace{0pt}
			elements\\
			Elevation Focus& 
			20&
			mm\\
			Element Height& 
			5&
			mm\\
			Element Width& 
			0.27&
			mm\\
			Kerf& 
			0.03&
			mm\vspace{2pt}\\
			\hline
		\end{tabular}
	\end{table}
	The center and sampling frequencies were set to 5.208 MHz and 20.832 MHz, respectively. It should be noted that due to the numerical precision of simulations in Field II, the initial sampling frequency was set to 104.16 MHz, and the simulated data was later downsampled by a factor of 5.
	All images were simulated using a full synthetic aperture scan, followed by synthesizing plane-wave images \cite{Rodriguez-Molares2015} with 384 columns from the acquired data and saved as RF data. We synthesized 100 randomly aberrated versions of each image according to the procedure elaborated in subsection \ref{sec:Methodology:PhaseAberrationImplementation:SimulatedAberration}. Although the non-aberrated version of images was not required for the proposed method, we opted to include them in the published dataset to enhance its comprehensiveness and facilitate the utilization of our data in a broader range of applications. This is because other methods may rely on non-aberrated images as a reference or ground truth. Fig. \ref{fig-simulated-dataset-samples} shows samples from the simulated dataset.
	
	\subsubsection{Experimental Phantom}
	\label{ssec:Datasets:ExperimentalPhantom}
	An L11-5v linear array transducer was operated using a Vantage 256 system (Verasonics, Kirkland, WA) to scan a multi-purpose multi-tissue ultrasound phantom (Model 040GSE, CIRS, Norfolk, VA). We acquired one scan of anechoic cylinders for evaluations and an additional 30 scans from other regions of the phantom for fine-tuning.
	In each acquisition, 51 single plane-wave images were captured, including one non-aberrated image (not for training and solely for visualization) and 50 randomly aberrated images utilizing pre-generated aberration profiles as elaborated in subsection \ref{sec:Methodology:PhaseAberrationImplementation:NumericalAberration}. To increase the frame rate, all 1550 required aberration profiles were randomly generated in advance and saved on the disk. Given the fixed position of both the probe and phantom and the sufficiently high frame rate, we assured that all 51 images belonged to the exact same region.
	
	\subsubsection{\textit{In-vivo}}
	\label{ssec:Datasets:InVivo}
	Two \textit{in-vivo} images acquired from the carotid artery of a volunteer with cross-sectional and longitudinal views were employed from the publicly available PICMUS dataset \cite{Liebgott2016}. Although this dataset was not explicitly designed for assessing aberration correction techniques, it was utilized due to the unavailability of other \textit{in-vivo} plane-wave images specifically acquired with aberrations as testing on a publicly available dataset allows other researchers to compare to our results.
	
	\subsection{Training}
	\label{ssec:Training}
	Inspired by Lehtinen \textit{et al.} \cite{Lehtinen2018}, the U-Net encoder-decoder CNN architecture \cite{Ronneberger2015} was employed to map beamformed RF data input to beamformed RF data target output, where both input and target output were distinct randomly aberrated versions of the same realization. The network was trained on 1800 simulated image sets for 5000 epochs, each set comprising 100 aberrated versions. In each epoch, a random pair of aberrated versions were mapped to each other. To optimize memory usage and accelerate the training process, we downsampled images laterally by a factor of 2, resulting in 192 columns for each image. Moreover, normalization was performed similar to \cite{Sharifzadeh2023a}, followed by applying the Yeo-Johnson power transformation \cite{Yeo2000}. A linear activation function was employed in the last layer, and the batch size was set to 32. We utilized Adam \cite{Kingma2015} with a zero weight decay as the optimizer. The learning rate was initially set to $10^{-3}$ and halved at epochs 500, 1000, 1500, and 4000.
	Fine-tuning on experimental images was performed with the same configurations by extending the training by an additional 20\% of the original epochs while utilizing a constant and substantially lower learning rate of $5\times10^{-5}$. To mitigate the impact of non-stationarity and attenuation in RF data training, we partitioned experimental images into three axial sections, each with a 3\% overlap, and fine-tuned a distinct network for each depth. When testing experimental images, we fed each image depth to its corresponding network and patched the outputs by blending the envelope of overlapping margins using weighted spatial averaging before displaying the final image.
	We implemented the method using PyTorch and trained all the models on two NVIDIA A100 GPUs in parallel.
	
	\subsection{Loss Function}
	\label{ssec:LossFunction}
	Let $\boldsymbol{S}, \boldsymbol{S'}, \boldsymbol{\hat{S}}$ $\in \mathbb{R}^{p\times q}$ represent input aberrated RF data, target output aberrated RF data, and network output, respectively. The aberration-to-aberration problem can be formulated as
	\begin{equation}
		\boldsymbol{\hat{S}} = f_{cnn}(\boldsymbol{S}, \boldsymbol{\theta}),
	\end{equation}
	\begin{equation}
		\boldsymbol{\theta^{\ast}} = \underset{\theta}{argmin} \; 	L(\boldsymbol{S'}, \boldsymbol{\hat{S}}),
	\end{equation}
	\noindent where $f_{cnn}: \mathbb{R}^{p\times q} \rightarrow \mathbb{R}^{p\times q}$ is the U-Net, $\boldsymbol{\theta}$ are the network's parameters, and during the training phase, an optimizer is utilized to find optimal parameters $\boldsymbol{\theta^{\ast}}$ that minimize the error, measured by a loss function $L$, between network's output $\boldsymbol{\hat{S}}$ and target output $\boldsymbol{S'}$. In this problem, input and target output were highly fluctuating aberrated RF data, which were randomly substituted at each epoch. We demonstrated in a pilot study in \ref{sec-results-pilotstudy} that the network encounters challenges in mapping pairs when comparing RF data directly using a conventional MSE loss defined as
	\begin{equation}
		L_{mse}(\boldsymbol{S'}, \boldsymbol{\hat{S}}) = \frac{1}{p\times q}||\boldsymbol{S'}-\boldsymbol{\hat{S}}||^2.
	\end{equation}
	On the other hand, we illustrated that training the network using the same loss function but on B-mode data leads to improved convergence, which can be attributed to the smoother loss landscape associated with B-mode data. Nonetheless, this improved convergence comes at the expense of discarding valuable information present in RF data. To leverage the benefits of both data types, we proposed an adaptive mixed loss function that gradually shifts from B-mode data to RF data as the training progresses,
	\begin{equation}
		\label{eq-adaptivemixedloss}
		\begin{aligned}
			L_{\text{adaptive\_mixed}}(\boldsymbol{S'}, \boldsymbol{\hat{S}}) = {} & (1-\alpha)L_{mse}(\mathcal{B}\{\boldsymbol{S'}\}, \mathcal{B}\{\boldsymbol{\hat{S}\}}) \\&+ \alpha L_{mse}(\boldsymbol{S'}, \boldsymbol{\hat{S}}),
		\end{aligned}
	\end{equation}

	\begin{equation}
		\begin{aligned}
			\alpha= \frac{\mathrm{current\ epoch\ number}}{\mathrm{total\ number\ of\ epochs}},
		\end{aligned}
	\end{equation}
	where $\mathcal{B}\{.\}$ denotes the log-compressed envelope data standardized by mean subtraction and division by its standard deviation.
	
	Our interpretation suggests that the proposed loss function guides the optimizer towards a correct solution by initially utilizing simpler data, before gradually incorporating more complex, fluctuating RF data to take full advantage of the richer information, like curriculum learning \cite{Bengio2009}. This helps to avoid getting stuck in local minima during the initial stages of the optimization.
	
	\subsection{Methods for Comparison}
	\label{ssec:MethodsforComparison}
	Among recent aberration correction methods, many either require multiple plane-wave transmissions \cite{Stahli2020, Bendjador2020} or utilize multistatic aperture data for synthetic focusing across all points \cite{Ali2022}. We compared the proposed method with two approaches applicable to single plane-wave images, enabling a fair comparison.
		\subsubsection{Beamsum}
		\label{ssec:MethodsforComparison:Beamsum}
		The beamsum method, which has recently demonstrated promising results in cardiac imaging \cite{Masoy2023}, estimates delay errors by maximizing normalized cross-correlation between individual channel signals and a common reference signal, known as the beamsum \cite{ODonnell1992}.
		In this method, after applying beamforming time delays, all channel signals are summed to form the reference signal. Subsequently, each channel signal is aligned with the reference one by maximizing their normalized cross-correlation. 
		A potential limitation arises from a relatively low correlation between individual channel signals and the beamsum, especially in plane-wave images with limited steering angles. To mitigate this, we averaged each channel signal with those from its $n$ adjacent channels before being compared with the beamsum. While this averaging might theoretically reduce the accuracy of the aberration profile estimation, it enhances overall performance in practical applications when the correlation is low.
		In this study, we heuristically set $n$ to 4 to achieve optimal performance. Additionally, to ensure a fair comparison, corrections using the beamsum method were only applied during reception, without iterative corrections during subsequent transmissions.
		
		\subsubsection{FXPF}
		\label{ssec:MethodsforComparison:FXPF}
		The FXPF method has proven effective in filtering out acoustic clutter and random noise \cite{Shin2017}, and its application has expanded to include mitigating noise induced by phase aberration \cite{Shin2018a}. Let us consider the received RF signal at time $t$ by element $n$ located at $x_n$ and denote its Fourier transform as 
	$RF_{n}(f)=\mathscr{F}\{RF(x_{n},t)\}$.
	The FXPF method establishes an autoregressive model of order $d$ across the RF channel signals received at transducer elements. Specifically, in the frequency domain and for each temporal frequency $f_0$, the method predicts a signal as a linear combination of the signals received by the preceding channels:
	\begin{equation}
		\label{eq-fxpf1}
		\begin{aligned}
			RF_{n+1}(f_0)=b_1RF_{n}(f_0)+b_2RF_{n-1}(f_0)+\\...+b_dRF_{n+1-d}(f_0).
		\end{aligned}
	\end{equation}
	Estimating coefficients $b$ from noisy data filters out non-conforming components based on the established model. Further details can be found in \cite{Shin2017, Shin2018a}.
	Although FXPF had been previously employed for focused images, we adapted this method for plane-wave images. The key adjustment involved applying apodization before using the method on the data to avoid image deterioration at shallow depths. This alteration was necessary due to significant variation in channel data across different elements at these depths, where signals from more distant elements are inaccurate and negatively affect the autoregressive model.
	
	In all experiments, the FXPF method was employed with an autoregressive model of order 2 and 3 iterations, determined to yield the optimal performance through a 6$\times$6 grid search, with each parameter ranging from 1 to 6. Consistent with the original study, we set a stability factor of 0.01 and a kernel size equivalent to one wavelength. As the implementation of this method was not publicly available, we took the initiative to publicly release our own implementation, to enhance the reproducibility of the reported results.
	
	\subsection{Quality Metrics}
	\label{ssec:QualityMetrics}
	To quantitatively measure the quality of reconstructed images, we calculated contrast, generalized contrast-to-noise ratio (gCNR) \cite{Rodriguez-Molares2020}, speckle signal-to-noise ratio (SNR), and full width at half maximum (FWHM) metrics for the test images:
	\begin{equation}
		Contrast=-20\log_\mathrm{10}(\frac{\mu_{t}}{\mu_{b}})\label{eq1},
	\end{equation}
	\begin{equation}
		SNR=\frac{\mu_{b}}{\sigma_{b}}\label{eq3},
	\end{equation}
	\begin{equation}
		\label{eq-gcnr}
		gCNR=1-\int_{-\infty}^{+\infty}{\min\limits_{\substack{x}}\{p_{t}(x),\ p_{b}(x)\}dx},
	\end{equation}
	where $t$ and $b$ stand for target and background regions, respectively, $\mu$ is the mean, and $\sigma$ is the standard deviation. In (\ref{eq-gcnr}), $x$ denotes the image value at any given pixel, and $p(x)$ is the probability density function of the values taken by pixels of a region. The gCNR ranges from 0 to 1, with a higher value indicating better contrast. To provide a fair comparison, all metrics were calculated on the envelope-detected image in the linear domain before applying the log-compression and its subsequent changes to the dynamic range.
	
	\section{Results}
	\subsection{Pilot Study}
	\label{sec-results-pilotstudy}
	\begin{figure}
		\includegraphics[width=0.975\linewidth]{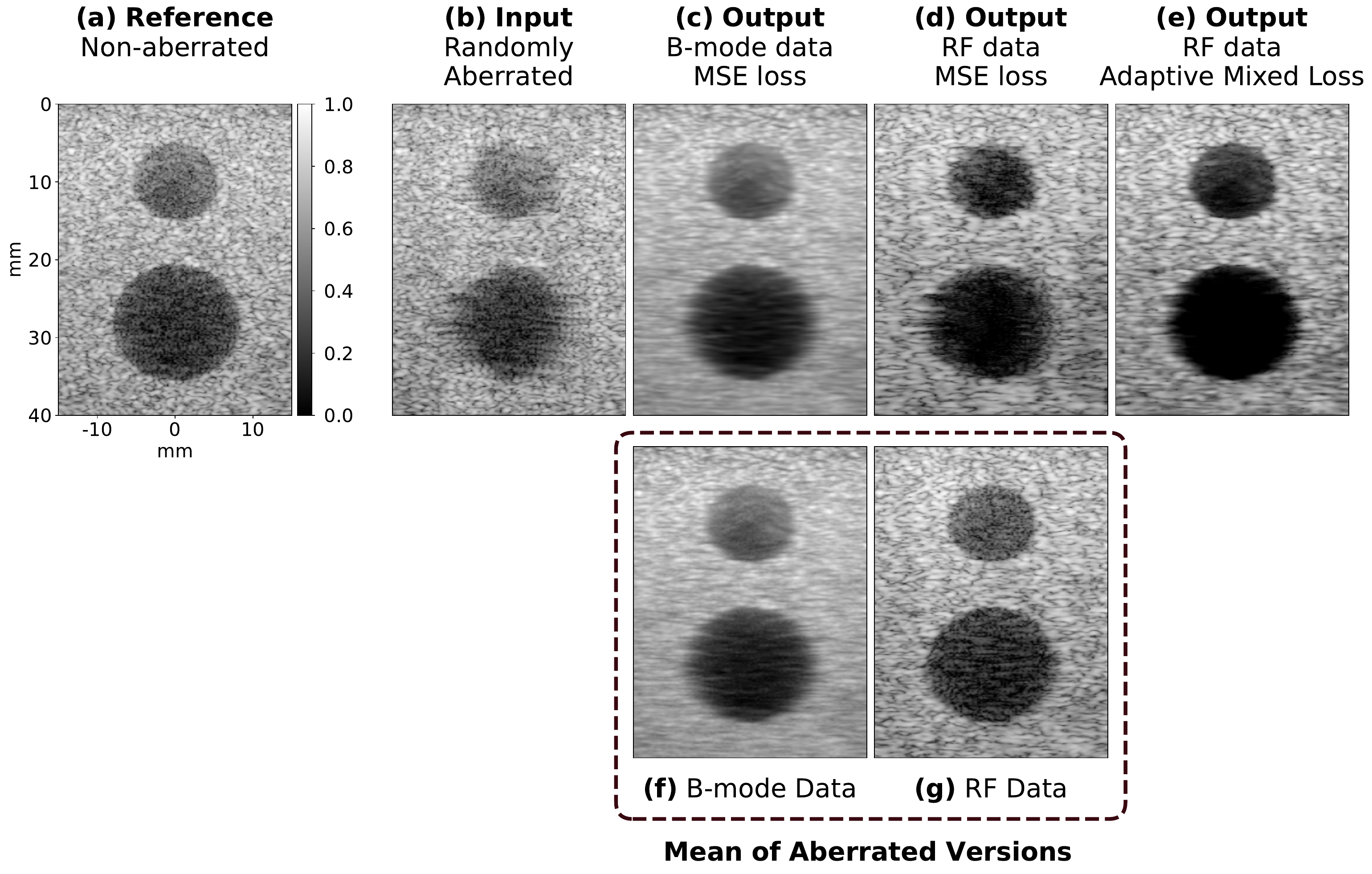}
		\caption{Training with different data types and loss functions in a pilot study. (a) The non-aberrated image, shown merely as a reference and not used for training. (b) The aberrated input image. The output of the network when it is trained on (c) B-mode data using the MSE loss function, (d) RF data using the MSE loss function, and (e) RF data using the proposed adaptive mixed loss function. Moreover, (f) and (g) show the mean of 99 aberrated versions that served as the training set in this pilot study, separately for B-mode and RF data. All images were normalized to their maximum intensity value and displayed on a 50 dB dynamic range.}
		\label{fig-compare-losses}
		\vspace{-5pt}
	\end{figure}
	To assess the performance of the proposed adaptive mixed loss function, we conducted an isolated pilot study using solely the simulated contrast test set described in \ref{ssec:Datasets:Simulated}. That set consisted of 100 aberrated versions of the same realization, where 99 versions served as the training set, and the remaining one version was used for testing in this pilot study. The network was trained using the configuration specified in Section \ref{ssec:Training}, in which during each epoch, each of the 99 versions was randomly mapped to another one. We trained three distinct networks, fed them with the test version, shown in Fig. \ref{fig-compare-losses}(b), and compared their outputs. The first network was trained using B-mode data, both as input and output, utilizing MSE loss. The resulting output is depicted in (c), where cyst boundaries were mostly recovered, but the image appears to be blurry compared to the non-aberrated (a) and aberrated (b) images. This blurring effect is consistent with the findings reported in \cite{Gobl2022}, where the objective was speckle filtering.
	To further illustrate the principles outlined in Section \ref{ssec:A2A} regarding the aberration-to-aberration approach, we averaged the 99 aberrated versions that served as the training set in this pilot study, separately for B-mode and RF data. The results are showcased in (f) and (g), which are aligned with findings in \cite{Laugier1990}. Interestingly, the network output in (c) closely resembles that of (f), indicating that the first network, trained with MSE loss, attempted to average the aberrated B-mode targets.
		However, the speckle pattern contains valuable information that can be utilized in applications such as elastography \cite{Tehrani2022}. Motivated by the richer information content present in RF data and aiming for a sharper output similar to the one shown in (g), we trained the second network using RF data as both input and output. As shown in (d), the network encountered challenges in mapping pairs of highly fluctuating aberrated RF data, which were randomly substituted at each epoch, leading to limitations in recovering cyst boundaries compared to the B-mode data scenario.
	
	Inspired by the results obtained from training with B-mode and RF data, we combined both approaches by training the third network using RF data as both input and output but employing the proposed adaptive mixed loss function. The proposed loss function gradually shifts from B-mode data to RF data as the training progresses toward convergence. As shown in (e), this approach exploited the advantages of the rich information within RF data and produced a sharper image compared to (c) while still retaining the ability to recover boundaries more efficiently compared to (d).
	The enhanced contrast compared to (a) and (g) is also elaborated upon in the Discussion section.
	Although further investigations are required, we believe that the advantages of the proposed loss function extend beyond the aberration correction task and can potentially improve the performance of other networks working with RF data in various tasks.
	
	\subsection{Main Study}
	\label{sec-results-mainstudy}
	\begin{figure*}
		\centering
		\includegraphics[width=0.992\linewidth]{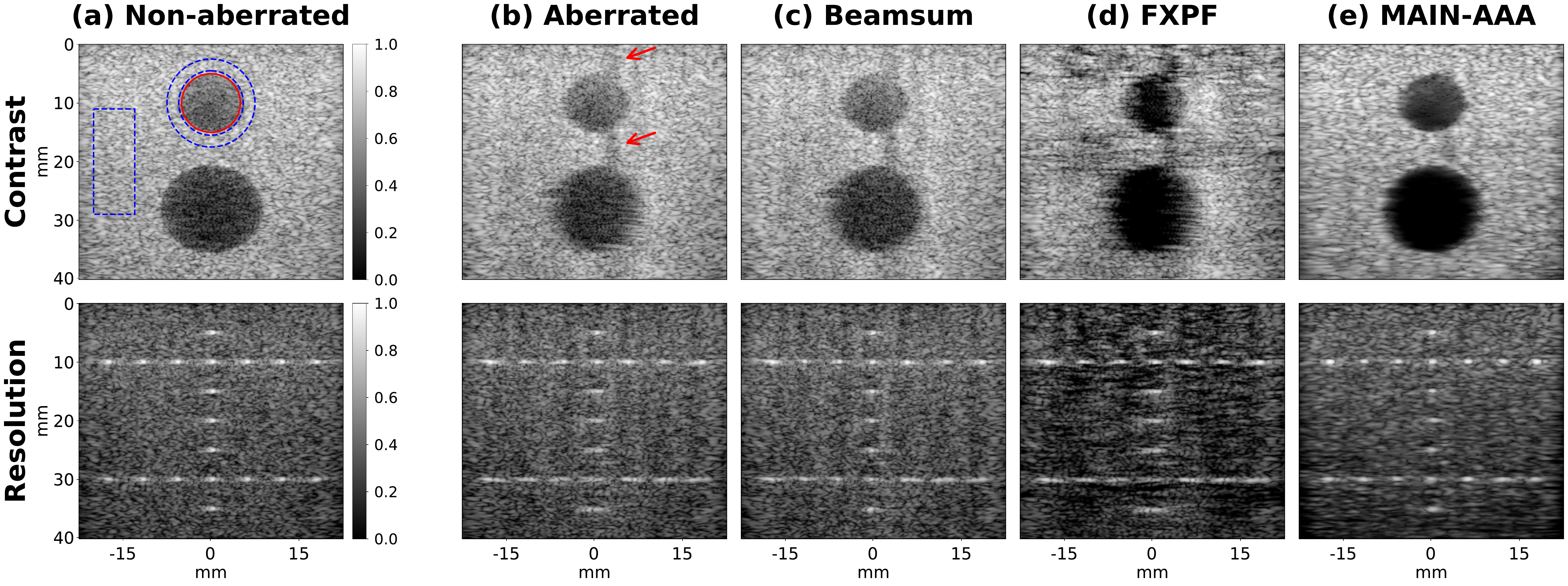}
		\vspace{-12pt}
		\caption{Simulated contrast and resolution test images. (a) The non-aberrated image, shown merely as a reference and not used for training. (b) A sample aberrated image reconstructed using DAS. (c) Beamsum output. (d) FXPF output. (e) MAIN-AAA output. All images were normalized to their maximum intensity value and displayed on a 50 dB dynamic range.}
		\label{fig-results-simulation}
	\end{figure*}
	Based on the findings from the pilot study, we chose the adaptive mixed loss function and utilized it for the subsequent experiments presented in this paper. In the main study, we trained the network using 1800 simulated image sets and evaluated its performance on two contrast and resolution test sets, each including 100 aberrated versions of the corresponding image. One such aberrated version, reconstructed using the conventional delay-and-sum (DAS), is shown in Fig. \ref{fig-results-simulation}(b) for each test image, followed by the resulting outputs of the beamsum, FXPF, and the proposed method. Note that in contrast to the pilot study, the network, in this case, was trained on images similar to those depicted in Fig. \ref{fig-simulated-dataset-samples} and had never seen, for instance, a perfectly circular cyst during the training phase.
	
	\begin{figure*}
		\centering
		\includegraphics[width=0.994\linewidth]{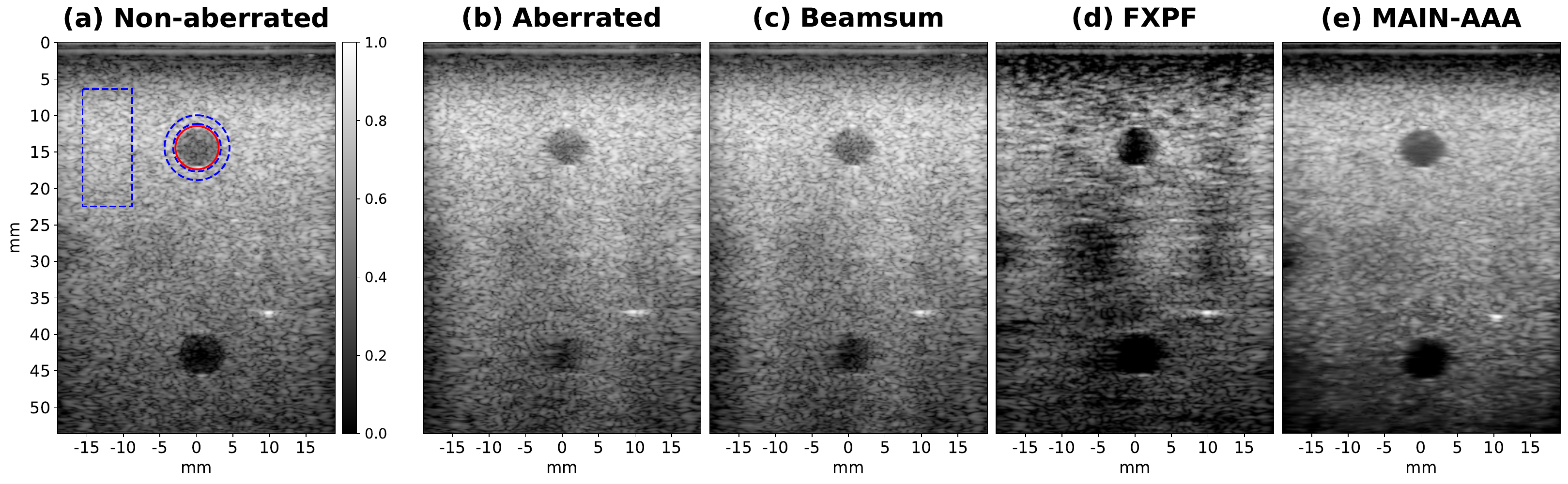}
		\vspace{-12pt}
		\caption{Experimental phantom results with quasi-physical aberrations. (a) The non-aberrated image, shown merely as a reference, and not used for training. (b) A sample aberrated image reconstructed using DAS. (c) Beamsum output. (d) FXPF output. (e) MAIN-AAA output. All images were normalized to their maximum intensity value and displayed on a 60 dB dynamic range.}
		\label{fig-results-numerical_phantom}
	\end{figure*}

\begin{figure}
		\centering
		\includegraphics[width=0.996\linewidth]{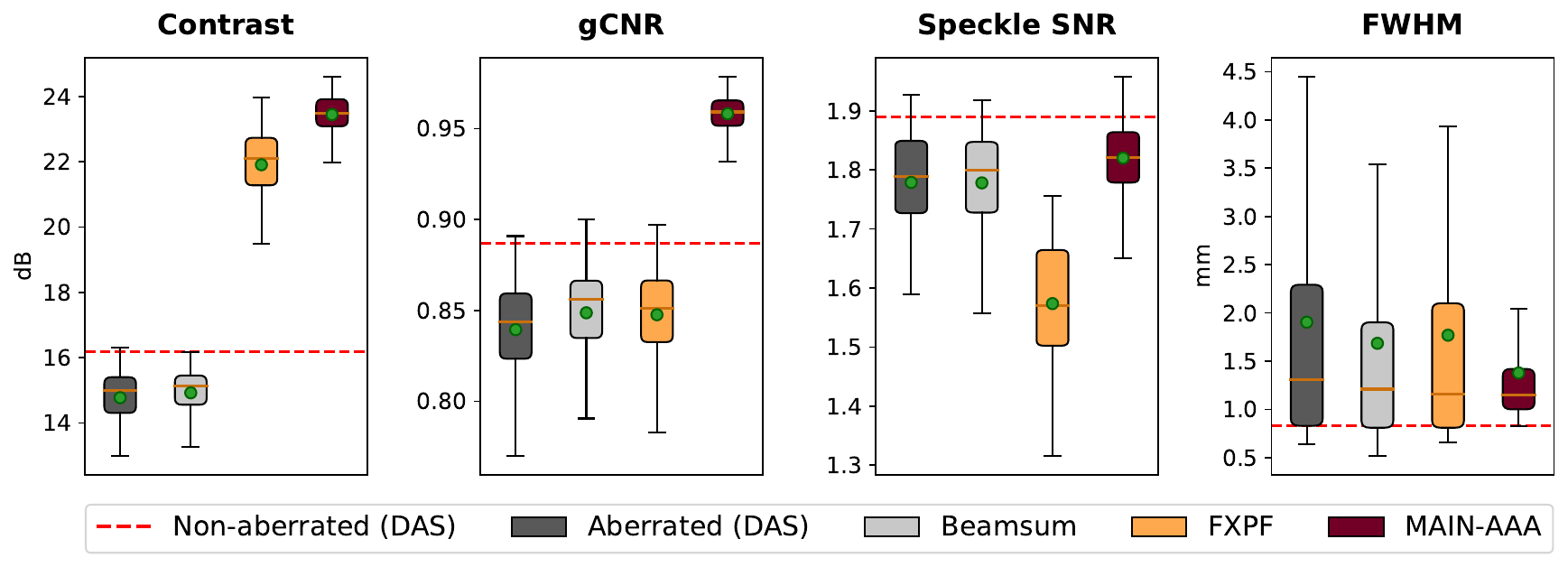}
		\vspace{-15pt}
		\caption{Quality metrics computed across the simulated test images. Contrast (dB), gCNR, and speckle SNR metrics computed across the contrast test set, with the FWHM metric obtained across the resolution test set. The green circle and orange horizontal line represent the mean and median, respectively.}
		\label{fig-boxplots-simulation}
		\vspace{3pt}
	\end{figure}

	To perform a quantitative evaluation, we calculated the average values of contrast and gCNR across the top and bottom anechoic cysts, as well as  the speckle SNR in the contrast test image. Although target and background regions, used for calculating metrics, were chosen similarly for both cysts, they are depicted only for the top cyst in the non-aberrated image in Fig. \ref{fig-results-simulation}(a) for brevity. For these metrics, the target region was inside a concentric circle with the same radius as that of the cyst (solid red circle). For contrast and gCNR, the background was the region between two concentric circles with radii of 1.1 and 1.5 times the cyst radius (dashed blue circles), while for speckle SNR, it was inside a rectangle far from the cyst (dashed blue rectangle). Additionally, to evaluate resolution, FWHM was measured for 19 point targets within the resolution test image in the lateral direction. To isolate the FWHM values of each point target from its adjacent ones, we confined the lateral profile to a 4 mm span on either side. The results were obtained for 100 aberrated versions of each test image and are shown in Fig. \ref{fig-boxplots-simulation}.

\begin{figure}
		\centering
		\includegraphics[width=0.996\linewidth]{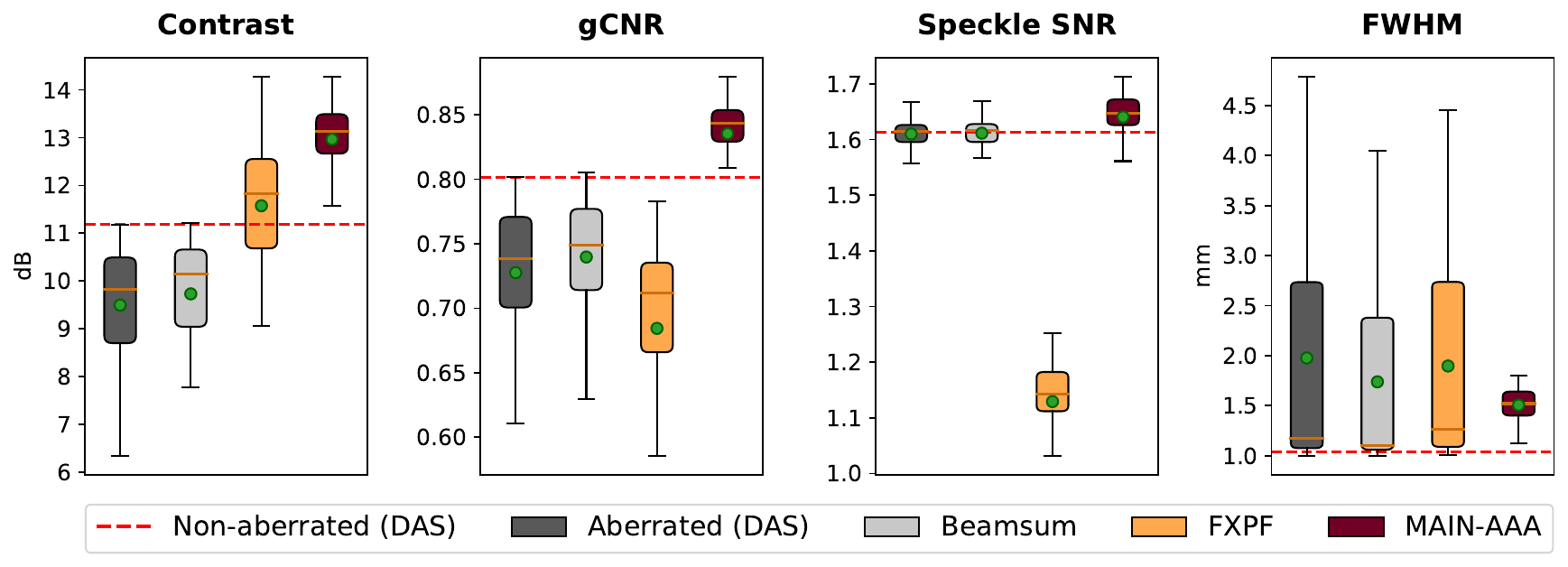}
		\vspace{-15pt}
		\caption{Quality metrics computed across the test images from the experimental phantom with quasi-physical aberrations. Contrast metrics were determined using the top and bottom cysts, while the resolution metric was based on the point target positioned at a depth of 37 mm. The green circle and orange horizontal line represent the mean and median, respectively.}
		\label{fig-boxplots-numerical-phantom}
		\vspace{-5pt}
	\end{figure}

	Fig. \ref{fig-results-numerical_phantom} presents the results for one of the aberrated versions of the experimental phantom test image, which was acquired with quasi-physical aberrations as explained in \ref{sec:Methodology:PhaseAberrationImplementation:NumericalAberration}. The quality metrics were calculated similar to those for the simulated test sets. The top and bottom anechoic cysts were utilized for calculating contrast metrics, while the point target at a depth of 37 mm was employed for resolution metrics. These metrics were obtained for all 50 aberrated versions of the test set and presented in Fig. \ref{fig-boxplots-numerical-phantom}.
	
	\begin{figure*}[h!]
		\centering
		\includegraphics[width=0.98\linewidth]{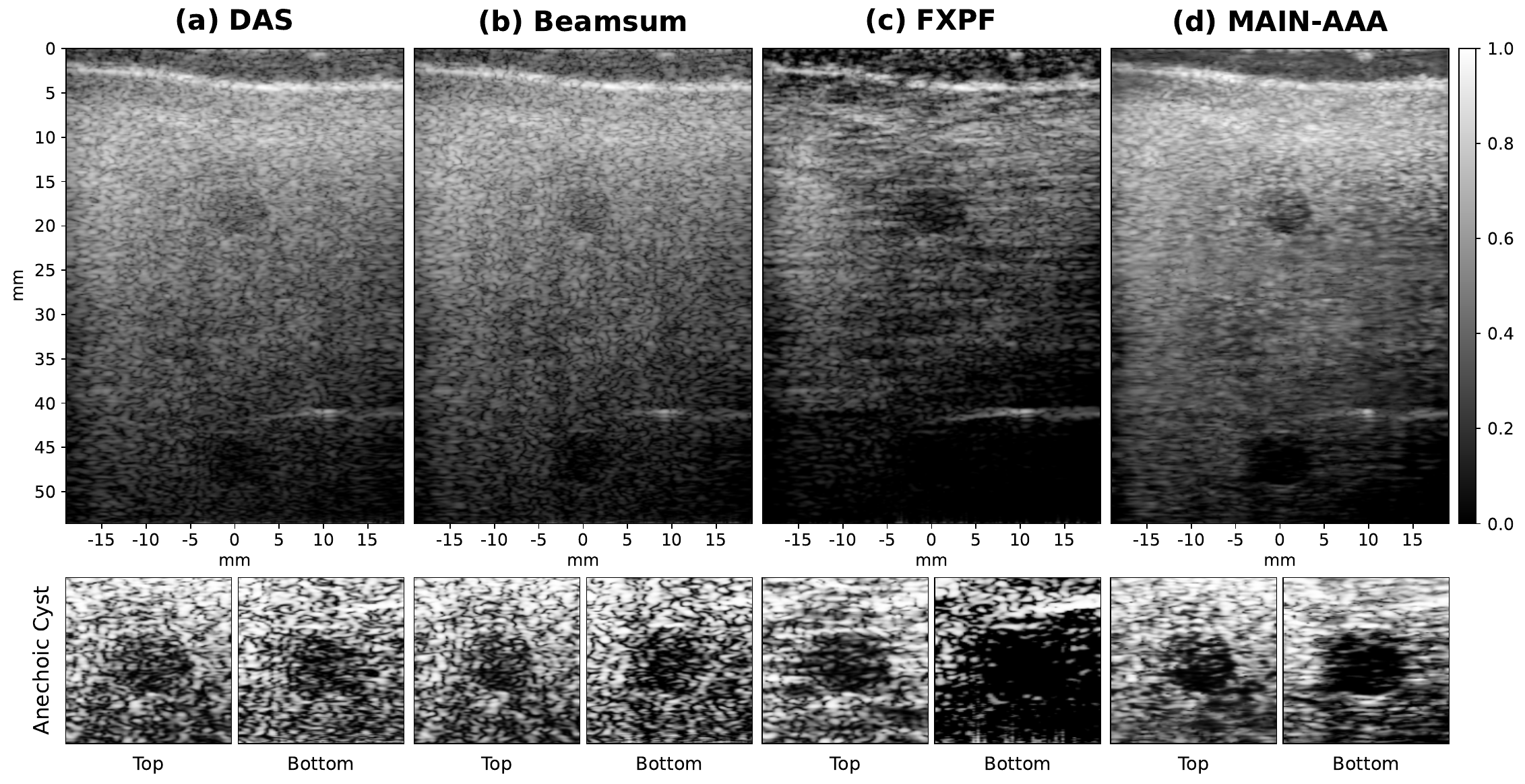}
		\vspace{-10pt}
		\caption{Experimental phantom aberrated using a physical aberrator layer. (a) DAS reconstruction. (b) Beamsum output. (c) FXPF output. (d) MAIN-AAA output. All images were normalized to their maximum intensity value and displayed on a 60 dB dynamic range. The second row shows cropped regions of interest (top and bottom anechoic cysts) corresponding to each image, where they were histogram-equalized to enhance visual comparability.}
		\label{fig-results-physical_phantom}
	\end{figure*}

	\begin{figure*}[h!]
		\centering
		\includegraphics[width=0.98\linewidth]{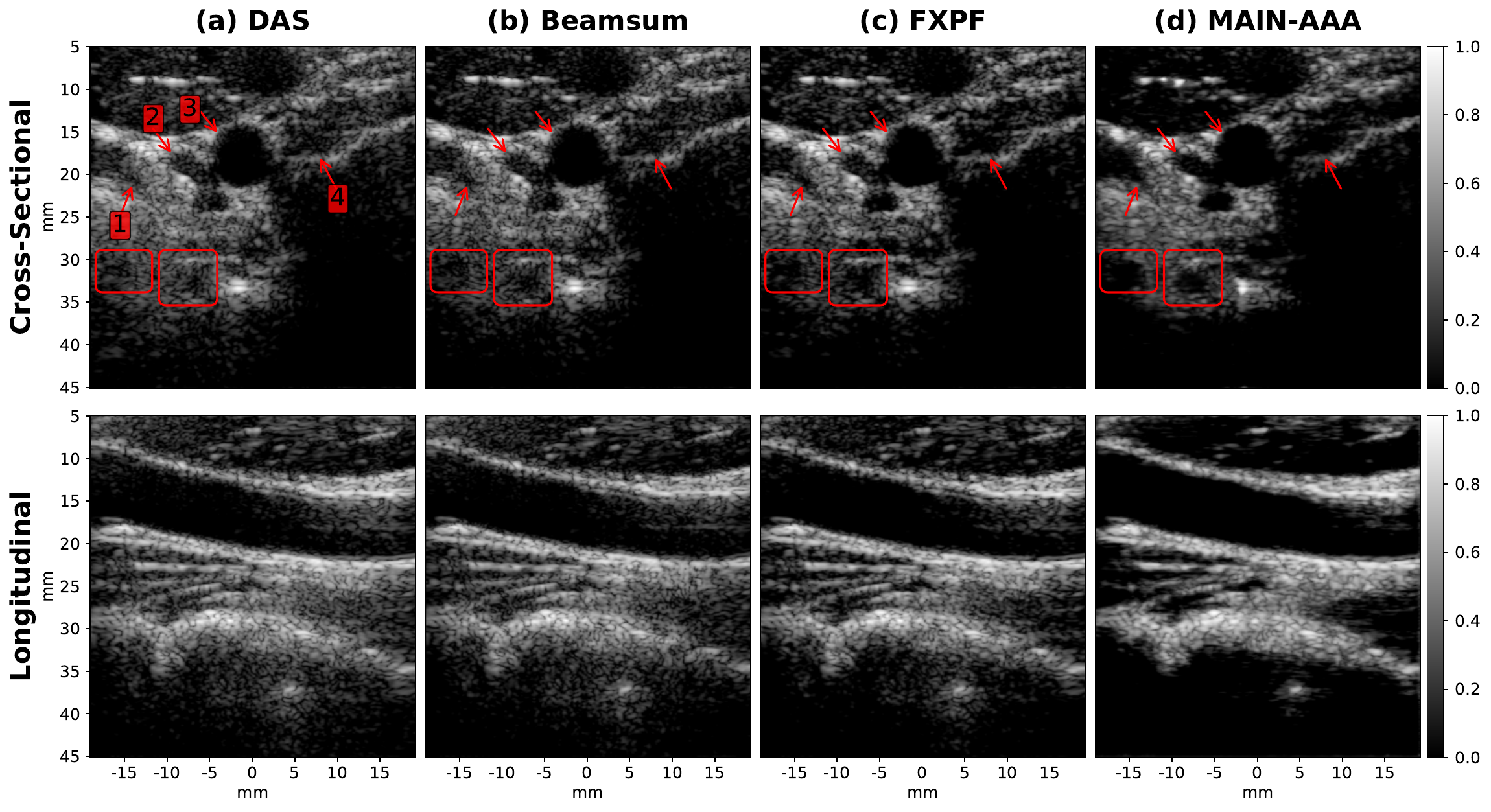}
		\vspace{-11pt}
		\caption{\textit{In-vivo} cross-sectional and longitudinal carotid artery images from the PICMUS dataset. (a) DAS reconstruction. (b) Beamsum output. (c) FXPF output. (d) MAIN-AAA output. All images were normalized to their maximum intensity value and displayed on a 50 dB dynamic range.}
		\vspace{-5pt}
		\label{fig-results-picmus}
	\end{figure*}
	
	Fig. \ref{fig-results-physical_phantom} presents the results for the experimental phantom aberrated with a physical aberrator. The first column displays the image reconstructed using conventional DAS, while the subsequent columns show the output images of the beamsum, FXPF, and proposed methods. To enhance visual comparability, the top and bottom cysts were cropped, then histogram-equalized, and displayed under their respective images. The results indicate that the proposed method outperformed the others in recovering cyst boundaries, especially the bottom one. Finally, we applied the methods to \textit{in-vivo} cross-sectional and longitudinal carotid artery images obtained from the PICMUS dataset. The results are shown in Fig. \ref{fig-results-picmus}, including annotations highlighting specific features, which will be further explained in the subsequent section.
	
	\section{Discussion}
	\label{sec:Discussion}
	In the pilot study, we conducted an experiment where we mapped different aberrated versions of the same image to each other to demonstrate the effectiveness of the proposed adaptive mixed loss function in correcting the phase aberration effect without over-smoothing the RF data and without requiring a non-aberrated ground truth. Notably, the results of this experiment also revealed an interesting finding. Specifically, it is evident in Fig. \ref{fig-compare-losses} that not only the phase aberration effect introduced in the input (b) is corrected in the output (e), but a higher contrast is achieved even compared to the non-aberrated reference image (a) and the averaged image (g).
	One possible explanation for this superior performance could be attributed to the ability of the network to leverage the RF data across the entire image and to average across all plausible explanations in order to output each region of the corrected version. By taking into account all the data points collectively, the network can make more informed decisions regarding each individual value during the reconstruction process. This can be analogized to the non-local means denoising algorithm \cite{Buades} in traditional image processing, which has been shown to outperform local filters in achieving higher performance. As a result, the network does not rely solely on RF data from local areas to correct the aberration but instead takes advantage of information from the entire dataset, resulting in an improved image compared to the reference image reconstructed using DAS.
	Another possible explanation is that the network develops the ability to effectively eliminate noise and clutter while randomly mapping one aberrated version to another. Since the noise and clutter are inconsistent across different aberrated versions, the network learns to disentangle them from the consistent tissue response by averaging plausible explanations.
	Interestingly, this finding aligns with the study by Jing \textit{et al.} \cite{Jing2020}, in which they proposed enhancing the spatial resolution of plane-wave images by introducing weak aberration into received data. They calculated the pixel-wise standard deviation of multiple aberrated versions and subtracted the result from the original image. Although our approach differs entirely from theirs, the concept of obtaining an enhanced image from its aberrated versions is similar and can explain the improvement over the reference image reconstructed using DAS.
	
	The main study involved training a general model on a dataset containing images similar to those presented in Fig. \ref{fig-simulated-dataset-samples} and evaluating its performance using contrast and resolution test sets, with sample images depicted in Fig. \ref{fig-results-simulation}. Red arrows in the aberrated contrast image (b) highlight the shadowing effect of the perturbed wavefront during transmission. The proposed method outperformed both beamsum and FXPF, which failed to detect and correct this effect. 
	As previously mentioned, the beamsum method was applied only during the reception, unable to compensate for this effect without iterative corrections during subsequent transmissions.
		Similarly, the FXPF method relies solely on the local signal information of a single image and eliminates components that do not conform to the autoregressive model across the echo signals received at the transducer elements. Nevertheless, in cases where all the echo signals experience a decrease in amplitude, the algorithm is unable to estimate a corrected signal with a higher amplitude. Instead, it tends to amplify the darkness of already dark regions, which may not necessarily correspond to anatomically relevant tissues, such as an anechoic cyst. The findings are consistent with the metrics reported in Fig. \ref{fig-boxplots-simulation}, indicating that while the FXPF algorithm improved contrast, it slightly impacted gCNR. Conversely, the proposed MAIN-AAA method enhanced contrast and achieved a higher gCNR of 0.96, which is substantially closer to the maximum value of 1.
	
	Similarly, we can observe in Fig. \ref{fig-results-numerical_phantom} that the proposed method recovered the size of the anechoic cyst at the bottom of the image more accurately, contrasting with the beamsum and FXPF methods which respectively led to an underestimation and overestimation of its size. As reported in Fig. \ref{fig-boxplots-numerical-phantom}, although the FXPF method improved the mean contrast of cysts, the mean gCNR actually decreased due to its aforementioned limitation. Conversely, the beamsum and proposed methods consistently enhanced both the contrast and gCNR metrics, with MAIN-AAA outperforming the other in both metrics. 
	In addition to the top and bottom anechoic cysts, this image contained four additional cysts positioned at a depth of 30 mm, arranged from left to right with contrast levels of -6 dB, -3 dB, +3 dB, and +6 dB relative to the background. It can be observed that, for instance, the -3 dB target was recovered with higher accuracy in terms of both its shape and contrast level, aligning with our prior knowledge that it was a hypoechoic cyst with a contrast level of -3 dB.
	
	In both simulation and phantom experiments, the FXPF method reduced speckle SNR, which aligns with the findings reported in \cite{Shin2018a}. As illustrated in Figs. \ref{fig-results-simulation} and \ref{fig-results-numerical_phantom}, this method increased the variance of the values within the blue rectangle while it preserved or even reduced their mean (darker region), thereby leading to a reduction in speckle SNR. In contrast, the proposed method consistently preserved the speckle SNR at a level approximately comparable to that of the aberrated image reconstructed using conventional DAS and the beamsum method. This preservation is deemed positive, given that the proposed method inherently operates by averaging all plausible explanations, thereby tending to smooth images. One of the objectives of introducing the adaptive mixed loss function was to prevent the network from over-smoothing images. While smoothing the speckle pattern could enhance the SNR, preserving it was desired for the proposed method.	
	
	Both simulation and phantom experiments demonstrated a relatively similar trend in resolution metrics. As shown in Fig. \ref{fig-boxplots-simulation} and Fig. \ref{fig-boxplots-numerical-phantom}, MAIN-AAA achieved the best mean FWHM, followed by the beamsum method. According to the mean values, the distributions of FWHM seemed more skewed in the simulation experiment compared to the phantom experiment because the experimental phantom images featured only one point target, whereas the simulated images contained 19 point targets distributed across different depths, with a less pronounced impact of aberration on shallower targets.
		While the mean FWHM across multiple data points can serve as a reliable measure for assessing resolution, it is worth noting that the phase aberration effect can also lead to artificially lower FWHM values. Such instances may occur when calculating the metric at the edge of the target or on a noisy profile,  often due to a displaced or entirely missing target. If a method mitigates the issue by partially recovering such a missing target, this recovery may contribute to increasing the FWHM value. This partially explains why the proposed method consistently yielded higher minimum FWHM values due to recovering more erroneous values. Another reason is its inherent averaging nature, which limits its ability to achieve resolutions equivalent to non-aberrated versions.
	
	Fig. \ref{fig-results-physical_phantom} shows a similar trend as observed in Figs. \ref{fig-results-simulation} and \ref{fig-results-numerical_phantom}, where the improvement in the bottom cyst is more prominent than in the top cyst. This observation can be attributed to the fact that the severity of the phase aberration effect, which needs to be corrected, is lower at shallower depths compared to deeper depths for two reasons. Firstly, perturbations in the wavefront escalate with propagation, leading to an increase in the aberration effect during transmission as the wavefront moves forward. Secondly, in plane-wave imaging, the reconstruction process uses a smaller aperture size $a$ at shallower depths and gradually increases it for deeper depths according to the $f\text{-}$number. Employing fewer neighboring elements during the reconstruction of lower depths can limit the aberration effect, especially when the aberration profile lacks abrupt changes, as a smaller segment of the aberration profile directly affects the reconstruction.
	
	In Fig. \ref{fig-results-picmus}, we compared the methods on cross-sectional and longitudinal views of the \textit{in-vivo} carotid artery image obtained from the PICMUS dataset.
	Despite the absence of intentional aberrations and any ground truth, we observed interesting results in these images. Several parts of the images were modified after applying the aberration correction methods, and we highlighted some of the notable alterations.
	In the cross-sectional view, the reconstructed image obtained by the DAS method exhibited indistinct boundaries for the right subclavian vein (arrow \#1). Although the FXPF method mitigated the issue, the proposed MAIN-AAA method achieved a higher level of boundary contrast, allowing differentiation of the vessel wall and its anechoic lumen.
	A similar trend was observed for the right jugular vein (arrow \#2) and right common carotid artery (arrow \#3), wherein the proposed method reconstructed images with superior tissue differentiation as evidenced by sharper boundaries and more distinct anatomical structures compared to other methods. Similarly, the proposed method demonstrated a higher contrast for the posterior edge of the right thyroid lobe (arrow \#4), thereby further enhancing the visual quality of the reconstructed image.
	Additionally, the MAIN-AAA method revealed ovoid-shaped structures within the two rectangles, which may represent the right subclavian artery (small rectangle) and right vertebral artery (large rectangle) seen cross-sectionally. These areas were barely visible using the DAS method, potentially because of their depth and size, as well as potential aberration induced by the highly anisotropic sternocleidomastoid muscle (the large oval at the top left of the figure).
	Although an aberration-free ground truth for this image is unavailable, this representation is supported by the similarity of the pattern within these rectangles in the DAS image to the aberrated anechoic cysts seen in the phantom images and the fact that both the beamsum and FXPF methods, which are not learning-based, identified them by attempting to enhance their contrast, while the proposed method produced substantially sharper boundaries for these tissues.
	In the longitudinal image, while both the beamsum and FXPF methods aimed to mitigate reverberation within the lumen of the common carotid artery at a depth of 17 mm, the proposed method outperformed them, resulting in a markedly sharper vessel wall contrast. A similar trend was observed for the muscle structure at depths ranging from 20 to 28 mm. Interestingly, regardless of the contrast improvement, both the beamsum and proposed methods slightly refined the bright point at a depth of 37 mm, reducing its dispersion. Although the absence of a ground truth complicates evaluating the accuracy of this modification, considerations such as the position of the bright point suggest that these methods have contributed to a more precise depiction.
	
	The capability of a network trained based on the near-field phase screen model to mitigate the noise induced by phase aberration in images such as the presented \textit{in-vivo} ones, which might be affected by distributed aberrations and do not necessarily adhere to the model, could be a subject of doubt. To address this concern, we can assume that the aberration at any given point within a medium is a consequence of variations in sound speed along the trajectory linking said point to each element of the aperture. Thus, regardless of the distribution of the aberrator, the variations in sound speed that contribute to the aberration of a particular point can still be approximated by a near-field phase screen. In other words, distributed aberrations throughout a heterogeneous medium can be characterized by multiple aberration profiles, each corresponding to a specific point.
	Chau \textit{et al.} \cite{Chau2019} developed a locally adaptive phase aberration correction method based on this assumption, estimating a local aberration profile at each point in the discretized image domain.
	\noindent While, theoretically, each point within the propagation medium may require a dedicated aberration profile, adopting the concept of finite-sized isoplanatic patches allows a single aberration profile to effectively model aberrations for all adjacent points within the same patch, which dramatically reduces the number of profiles required to characterize distributed aberrations across the medium.
	Consequently, we speculate that a network, trained and fine-tuned on over 180,000 unique aberration profiles, could learn from the presented variations and become capable of correcting distributed aberrations locally, even if they cannot be modeled by a single near-field phase screen.
	
	While the proposed method eliminates the need for ground truth data, a drawback is its reliance on averaging aberrated patterns to approximate the original tissue response. We investigated the incorporation of both low and high-frequency data, introducing an adaptive mixed loss function to facilitate the network's utilization of RF data and produce sharper outputs. Nevertheless, the underlying averaging principle inherently results in a smoother speckle pattern compared to methods that directly compensate for delay errors, such as beamsum.
	Furthermore, although modeling phase aberrations using near-field phase screens effectively mitigated noise resulting from distributed aberrations by addressing them locally, the performance is ultimately restricted by the aberration model within the training dataset. Future studies could explore the utilization of more complex models for phase aberrations, potentially leading to improved method efficacy. Additionally, we demonstrated the effectiveness of an aberration-to-aberration approach using single plane-wave images. Expanding on this work by incorporating compounded plane-wave images with additional steering angles \cite{Montaldo2009a} or exploring other imaging modes presents potential avenues for future research.
	
	\section{Conclusions}
	\label{sec:Conclusions}
	We proposed a novel DL-based approach for correcting phase aberration that eliminates the requirement for ground truths. We illustrated that a conventional loss function, such as MSE, is inadequate to achieve optimal performance and introduced an adaptive mixed loss function to train a network capable of mapping aberrated RF data to aberrated RF data. This approach permits training or fine-tuning using experimental images without prior assumptions about the presence or absence of aberration. Furthermore, we demonstrated the feasibility of obtaining the required data for this method using a programmable transducer and acquiring multiple aberrated versions of the same scene during a single scan. Apart from releasing the code for the proposed and the FXPF method, we also made available to the public a dataset containing more than a thousand sets of single plane-wave images stored as RF data, where each set comprises 100 aberrated versions of the same realization along with corresponding aberration profiles and the non-aberrated version, aiming to facilitate the advancement of DL-based methods for correcting phase aberration in ultrasound images.
	
	\section*{Acknowledgment}
	We acknowledge the support of the Government of Canada’s New Frontiers in Research Fund (NFRF), [NFRFE-2022-00295].
	
	\ifCLASSOPTIONcaptionsoff
	\newpage
	\fi
	
	\bibliographystyle{IEEEtran}

\end{document}